\documentclass[letterpaper, conference,10pt] {IEEEtran}
\usepackage[ left=0.64in, right=0.64in, top=0.75in, bottom=0.75in]{geometry}
\usepackage{amsmath}
\usepackage{amssymb}
\usepackage{bm}
\usepackage{color}
\usepackage{graphicx}

\usepackage[linesnumbered,ruled,lined]{algorithm2e}

\usepackage[outdir=./]{epstopdf}
\usepackage{caption}
\usepackage{lipsum}
\usepackage{cite}
\usepackage{multirow}
\usepackage{empheq}
\usepackage{fancyhdr}
\usepackage[caption=false, font=footnotesize]{subfig}

\SetKwInput{Input}{input}
\SetKwInput{Output}{output}

\IEEEoverridecommandlockouts
\begin{document}
	\title{
		Rate-Splitting for Multi-User Multi-Antenna Wireless Information and Power Transfer}

		\author{
			\\[-4.5ex]	
	
				\IEEEauthorblockN{Yijie Mao,  Bruno Clerckx and Victor O.K. Li }
	\IEEEauthorblockA{$^*$The University of Hong Kong, Hong Kong, China,
		$^\dagger$Imperial College London, United Kingdom\\
		Email: $^*$\{maoyijie, vli\}@eee.hku.hk, $^\dagger$b.clerckx@imperial.ac.uk}

	\\[-9ex]
	\thanks{This work is partially supported by the U.K. Engineering and Physical
		Sciences Research Council (EPSRC) under grant EP/N015312/1.}}	

\maketitle

\thispagestyle{empty}
\pagestyle{empty}

\begin{abstract}
In a multi-user multi-antenna Simultaneous Wireless Information and Power Transfer (SWIPT) network, the transmitter  sends information to the Information Receivers (IRs) and  energy to Energy Receivers (ERs) concurrently. A conventional approach is based on Multi-User Linear Precoding (MU--LP) where each IR directly decodes the intended stream by fully treating the interference from other IRs and ERs as noise. In this paper, we investigate the application of linearly-precoded Rate-Splitting (RS) in Multiple Input Single Output (MISO) SWIPT Broadcast Channel (BC). By splitting the messages of IRs into private and common parts and encoding the common parts into a common stream decoded by all IRs, RS manages the interference dynamically.
The precoders are designed such that the Weighted Sum Rate (WSR) of  IRs is maximized under the total transmit power constraint and the sum energy constraint for ERs. 
Numerical results show that the proposed RS-assisted strategy provides a better rate-energy tradeoff in MISO SWIPT BC.  Under a sum energy constraint of  ERs, RS-assisted strategy achieves better WSR performance of  IRs than MU--LP and NOMA in a wide range of IR and ER deployments. 
Hence, we draw the conclusion that RS is superior for downlink SWIPT networks.
\end{abstract}


\vspace{-2mm}
\section{Introduction}
\vspace{-1mm}

\par In recent years, linearly-precoded Rate-Splitting (RS) has been recognized as a promising transmission strategy to enhance rate, robustness and Quality of Service (QoS)  for future generations of wireless communication systems. Inspired by the Han-Kobayashi scheme for the two-user interference channel \cite{TeHan1981}, in RS,  the message of each receiver is split into a common part and a private part at the transmitter \cite{RS2016hamdi,Mingbo2016,RSintro16bruno,hamdi2017bruno,mao2018EE,mao2018rate}. The common parts of all the receivers are jointly encoded into a common stream required to be decoded by all the receivers while the private parts are independently encoded into private streams for the corresponding receivers only. All the streams are linearly precoded and simultaneously transmitted to the receivers.  By allowing each receiver to first decode the common stream and use Successive Interference Cancellation (SIC) to remove the common stream before decoding the intended private streams, receivers are enabled to partially decode the interference and partially treat the remaining interference as noise.  A more general framework of RS, namely Rate-Splitting Multiple Access (RSMA), is proposed in \cite{mao2017rate}. RSMA has been shown to outperform Multi-User Linear Precoding (MU--LP) where each receiver directly decodes the intended message by fully treating the interference as noise and power-domain Non-Orthogonal Multiple Access (NOMA) (simply referred to as NOMA in the sequel)  where 
Superposition Coding (SC) and SIC are enabled respectively at the transmitter and receivers (SC--SIC) such that receivers with stronger channel strength are required to decode the messages of the receivers with weaker channel strength. However, most existing works on RS only consider Wireless Information Transfer (WIT) via Radio-Frequency (RF). 

\par As RF signals carry not only information but also energy, wireless transmission can be used not only for WIT but also for Wireless Power Transfer (WPT) where  Energy Receivers (ERs) are enabled to harvest energy from RF \cite{YongZheng2017WPT}. 
A unified approach to study WIT and WPT is Simultaneous Wireless Information and Power Transfer (SWIPT), which enables one to simultaneously transmit information and power to  Information Receivers (IRs) and   ERs, respectively\footnote{ ERs and IRs can be co-located (where ER and IR are the same device that is simultaneously receiving information and harvesting energy) or separated (where ER and IR are different devices) \cite{bruno2019swipt}. In this work, we only consider separated ERs and  IRs.}\cite{bruno2019swipt,RuiZhang2013SWIPT}. 
In multi-user multi-antenna SWIPT networks,   the fundamental tradeoff between rate-energy has become the  critical criterion for the precoder design at the transmitter. Efficient precoder design has been studied in the literature with different objectives, such as maximizing the Weighted Sum Rate (WSR) of  IRs  \cite{RuiZhang2013SWIPT, Luo2015SWIPT}, maximizing the harvested energy of ERs \cite{Xujie2014SWIPT} and maximizing the system energy efficiency \cite{Sun2014SWIPT,Jang2018SWIPT}.  All of the above works consider the use of the MU--LP strategy. Only the recent work \cite{Su2019SWIPT} investigates RS in Multiple Input Single Output (MISO) SWIPT Interference Channel (IC) and shows the robustness improvement of  RS over MU--LP with co-located ERs and IRs. To the best of our knowledge, the benefits of RS in MISO SWIPT Broadcast Channel (BC) has not been investigated yet. 

\par In this work, motivated by the benefits of RS in WIT and MISO SWIPT IC, we initiate the study of RS in MISO SWIPT BC. At the transmitter, the messages of each IR is split into a common part and a private part. The common parts of all the IRs are jointly encoded into a common stream while the private parts are independently encoded into private streams.  Each  encoded stream of  IRs as well as each ER is assigned with one dedicated transmission beam. At receiver sides, each IR is required to decode the common stream and use SIC to remove it before decoding the intended private stream while each ER directly harvests energy. Based on the proposed RS-assisted SWIPT model, we focus on  the precoder design of IRs and ERs at the transmitter by investigating the problem of  maximizing the WSR of the IRs subject to the minimum harvested energy constraint of ERs and total transmit power constraint. 
We propose a Weighted Minimum Mean Square Error (WMMSE) and Successive Convex Approximation (SCA) based algorithm to efficiently solve the problem. 
We demonstrate in the numerical results that the rate region of IRs in RS-assisted SWIPT is always equal to or larger than that of the existing MU--LP and SC--SIC assisted SWIPT for a given sum energy constraint of ERs. Specifically, in the scenario of two IRs and one ER, RS-assisted SWIPT achieves a larger rate-energy region than MU--LP when the energy requirement is close to the maximum value,  the IRs are orthogonal and the ER is close to neither IRs.  The benefit originates from the presence of the common stream that is not only used to transmit information to IRs but also carries power to the ERs.

\vspace{-0.5mm}
\section{System model and problem formulation}
\vspace{-0.5mm}
\label{sec: system model}
In this section, the system model of the proposed RS-assisted MISO SWIPT BC with separated ERs and IRs is specified and the WSR maximization problem is formulated.
\par Consider a downlink multi-user multi-antenna SWIPT system with one Base Station (BC) equipped with  $N_t$ antennas serving $K$ single-antenna IRs indexed by  $\mathcal{K}=\{1,\dots,K\}$  and $J$ single-antenna ERs indexed by $\mathcal{J}=\{1,\dots,J\}$. IRs and ERs are respectively implemented with Information Decoding (ID) and Energy Harvesting (EH). 
RS  is enabled at the BS for the information transmission of IRs. The messages $W_k$ of IR-$k$ is split into a common part $W_{c,k}$ and a private part $W_{p,k}$, $\forall k\in\mathcal{K}$. The common parts of all IRs $\{W_{c,1},\ldots,W_{c,K}\}$ are jointly encoded into the common stream $s_c^{\mathrm{ID}}$ while the private parts are independently encoded into the private streams $\{s_1^{\mathrm{ID}},\ldots,s_K^{\mathrm{ID}}\}$. The set of streams for IRs $\mathbf{s}^{\mathrm{ID}}=[s_c^{\mathrm{ID}},s_1^{\mathrm{ID}},\ldots,s_K^{\mathrm{ID}}]^T\in\mathbb{C}^{K+1}$ are linearly precoded using the precoder $\mathbf{P}=[\mathbf{p}_c,\mathbf{p}_1,\ldots,\mathbf{p}_K]$, where  $\mathbf{p}_c\in\mathbb{C}^{N_t\times1}$ is the precoder for the common stream. 
The energy signal $s_j^{\mathrm{EH}}$ of ER-$j$ carries no information. It can be any arbitrary random signal  provided that its power spectral density satisfies certain regulations on microwave radiation\footnote{This holds only under the linear model of the harvester assumed in this work. In practice, due to the harvester nonlinearity, this would not hold \cite{Bruno2016WPT,YongZheng2017WPT}.}. The set of streams for ERs $\mathbf{s}^{\mathrm{EH}}=[s_1^{\mathrm{EH}},\ldots,s_J^{\mathrm{EH}}]^T\in\mathbb{C}^{J\times 1}$ is linearly precoded at the transmitter using the precoder $\mathbf{F}=[\mathbf{f}_1,\ldots,\mathbf{f}_J]$, where $\mathbf{f}_j\in\mathbb{C}^{N_t\times1}$ is the precoder of the energy signal for ER-$j$. 
The transmit information-bearing signal $\mathbf{x}^{\mathrm{ID}}=\mathbf{P}\mathbf{{s}}^{\mathrm{ID}}$ is superposed with the energy-carrying signal  $\mathbf{x}^{\mathrm{EH}}
=\mathbf{F}\mathbf{{s}}^{\mathrm{EH}}$. The resulting transmit signal is given by
\begin{equation}
\begin{aligned}
\mathbf{x}=\mathbf{x}^{\mathrm{ID}}+\mathbf{x}^{\mathrm{EH}}
={\mathbf{p}_{c}s_{c}^{\mathrm{ID}}}+\sum_{k\in\mathcal{K}}\mathbf{p}_{k}s_{k}^{\mathrm{ID}}+\sum_{j\in\mathcal{J}}\mathbf{f}_{j}s_{j}^{\mathrm{EH}}.
\end{aligned}
\vspace{-2mm}
\end{equation}
Under the assumption that $\mathbb{E}\{\mathbf{{s}}^{\mathrm{ID}}(\mathbf{{s}}^{\mathrm{ID}})^H\}=\mathbf{I}$ and $\mathbb{E}\{\mathbf{{s}}^{\mathrm{EH}}(\mathbf{{s}}^{\mathrm{EH}})^H\}=\mathbf{I}$,
 the transmit power is constrained by  $\mathrm{tr}(\mathbf{P}\mathbf{P}^{H})+\mathrm{tr}(\mathbf{F}\mathbf{F}^{H})\leq P_{t}$. $P_t$ is the available  transmit power.

\par The respective signals received at IR-$k$ and ER-$j$  are 
\begin{equation}
\begin{aligned}
&y_{k}^{\mathrm{ID}}=\mathbf{{h}}_{k}^{H}\mathbf{{x}}+n_{k}^{\mathrm{ID}},\forall k\in\mathcal{K},\\
&y_{j}^{\mathrm{EH}}=\mathbf{{g}}_{j}^{H}\mathbf{{x}}+n_{j}^{\mathrm{EH}},\forall j\in\mathcal{J},
\end{aligned}
\end{equation}
where $\mathbf{{h}}_{k},\mathbf{{g}}_{j}\in\mathbb{C}^{N_{t}\times1}$ are the corresponding channel from  BS to IR-$k$ and  the channel from  BS to ER-$j$.  $n_{k}^{\mathrm{ID}},n_{j}^{\mathrm{EH}}$ are the respective Additive White Gaussian Noises (AWGNs) received at IR-$k$ and ER-$j$ with zero mean and unit variance. 
The transmit SNR is equal to  $P_t$. 
Perfect channel state information is assumed at the transmitter and the receivers. 

\par The energy precoder $\mathbf{F}$ carries no information. It is assumed to be perfectly known at  the transmitter and IRs. Hence, IRs are able to remove the interference caused by energy signals from $y_k$ before decoding the intended information signals.  Following the decoding order in the literatures of RS \cite{RS2016hamdi,hamdi2017bruno,mao2017rate}, each IR first decodes the common stream by fully treating all the private streams as interference before decoding the intended private stream.   The Signal-to-Interference-plus-Noise Ratio (SINR) of decoding the common stream $s_c^{\mathrm{ID}}$ at IR-$k$ is
	\vspace{-1.5mm}
\begin{equation}
\label{eq: common sinr}
\gamma_{c,k}(\mathbf{P})=\frac{\left|\mathbf{{h}}_{k}^{H}\mathbf{{p}}_{c}\right|^{2}}{\sum_{j\in\mathcal{K}}\left|\mathbf{{h}}_{k}^{H}\mathbf{{p}}_{j}\right|^{2}+1}, \forall k\in\mathcal{K}.
	\vspace{-1.5mm}
\end{equation}
After successfully decoding $s_{c}^{\mathrm{ID}}$ and removing its contribution from $y_{k}$, IR-$k$ decodes the intended private stream $s_k^{\mathrm{ID}}$ by treating the interference from other IRs as noise. The SINR of decoding the private stream $s_k^{\mathrm{ID}}$ at IR-$k$ is
	\vspace{-1.5mm}
\begin{equation}
\label{eq: private sinr}
\gamma_{k}(\mathbf{P})=\frac{\left|\mathbf{{h}}_{k}^{H}\mathbf{{p}}_{k}\right|^{2}}{\sum_{j\in\mathcal{K},j\neq k}\left|\mathbf{{h}}_{k}^{H}\mathbf{{p}}_{j}\right|^{2}+1}, \forall k\in\mathcal{K}.
	\vspace{-1.5mm}
\end{equation}
The corresponding achievable rates\footnote{We here assume conventional Gaussian signaling because we consider the linear model of the energy harvester \cite{bruno2019swipt}. With nonlinearity, such signaling would be suboptimal \cite{bruno2019swipt}.} of $s_{c}^{\mathrm{ID}}$ and $s_{k}^{\mathrm{ID}}$ at IR-$k$ are
$
\label{eq: individual rate}
R_{c,k}(\mathbf{P})=\log_{2}\left(1+\gamma_{c,k}(\mathbf{P})\right)\, \textrm{and}\,\,
R_{k}(\mathbf{P})=\log_{2}\left(1+\gamma_{k}(\mathbf{P})\right).
$
$s_c^{\mathrm{ID}}$  is required to be decoded by all IRs. To guarantee that all IRs can successfully decode the common stream $s_c^{\mathrm{ID}}$, the achievable  rate of $s_c^{\mathrm{ID}}$ shall not exceed 
$
\label{eq: common rate}
R_{c}(\mathbf{P})=\min\left\{ R_{1,c}(\mathbf{P}),\ldots,R_{K,c}(\mathbf{P})\right\}
$ .  
As $R_{c}(\mathbf{P})$ is shared by $K$ IRs, we have $\sum_{k\in\mathcal{K}}C_{k}=R_{c}(\mathbf{P}) $ where $C_k$  is the introduced variable representing the portion of common rate $R_{c}(\mathbf{P})$ transmitting $W_{c,k}$. 
The total achievable rate of IR-$k$ contains the portion of common rate transmitting $W_{c,k}$ and  private rate transmitting $W_{p,k}$, which is given by
$
R_{k, tot} (\mathbf{P})=C_{k}+R_k(\mathbf{P})
$.

 The energy carried by both information and energy precoders is harvested at each ER. The resulting harvested energy at ER-$j$ is proportional to the total power received  \cite{Xujie2014SWIPT}, which is
 	\vspace{-1.5mm}
 \begin{equation}
 \resizebox{0.99\hsize}{!}{$	Q_j=\zeta\left(\left| \mathbf{g}_j^H\mathbf{p}_c \right|^2 +\sum_{k\in\mathcal{K}}\left| \mathbf{g}_j^H\mathbf{p}_k \right|^2+\sum_{j'\in\mathcal{J}}\left| \mathbf{g}_j^H\mathbf{f}_{j'} \right|^2\right),\forall j\in\mathcal{J}$}
 	\vspace{-1.5mm}
 \end{equation}
where $0\leq\zeta\leq1$ is the energy harvesting efficiency. Without loss of generality, we assume $\zeta=1$ in the rest of the paper\footnote{As a first attempt to identify the gain of RS, we here use the linear model though it is somewhat oversimplified and unrealistic \cite{YongZheng2017WPT,bruno2019swipt}. The study of the nonlinear models \cite{YongZheng2017WPT,bruno2019swipt} on RS SWIPT is left for future studies.}
.

\par In this work, we aim at achieving the optimal rate-energy tradeoff by maximizing the WSR of all IRs under the total transmit power constraint and the sum harvested energy constraint of ERs. Denote the weight allocated to IR-$k$ as $u_k$. The formulated optimization problem  is given by
	\vspace{-1.5mm}
\begin{subequations}
	\label{eq: RE prob}
	\begin{align}
	\max_{\mathbf{{P}},\mathbf{{F}},\mathbf{c}}&\sum_{k\in\mathcal{K}}u_k\left(C_k+R_{k}(\mathbf{P})\right)  \\
	\mbox{s.t.}\quad
	& \sum_{k\in\mathcal{K}}C_{k}\leq R_{c,k}(\mathbf{P}),\forall k\in\mathcal{K} \label{const: 1}\\
	&\sum_{j\in\mathcal{J}}Q_j\geq E^{th} \label{const: 3}\\	&\text{tr}(\mathbf{P}\mathbf{P}^{H})+\mathrm{tr}(\mathbf{F}\mathbf{F}^{H})\leq P_{t} \label{const: 4}\\
	&\mathbf{c}\geq\mathbf{0}
		\vspace{-2mm}
	\end{align}	
\end{subequations}
where  $\mathbf{c}=[C_1,\ldots,C_K]$ is the common rate vector.   Constraint (\ref{const: 1}) ensures that each IR is able to  decode the common stream. 
Constraint (\ref{const: 3}) is the  harvested energy  constraint. $E^{th}$ is the minimum  harvested energy requirement of ERs.

\par RS reduces to MU--LP by allocating no power to $s_{c}$. In a 2-IR setup, RS reduces to SC--SIC by forcing one user,
say IR-1, to fully decode the message of the other user,
say IR-2. This is achieved by allocating no power to $s_2$,
encoding $W_1$ into $s_1$ and encoding $W_2$ into $s_{c}$ \cite{mao2017rate}.

\section{Proposed WMMSE-SCA Algorithms}
\label{sec: RS}
The R-E tradeoff problem (\ref{eq: RE prob}) is non-convex. In this section, we introduce the use of WMMSE and SCA based algorithm.  

The common stream $s_c^{\mathrm{ID}}$ is firstly decoded from the received signal at IR-$k$ via an equalizer $g_{c,k}$. Once $s_c^{\mathrm{ID}}$ is successfully decoded and removed from the received signal, the private stream $s_k^{\mathrm{ID}}$ is decoded by using the equalizer $g_k$. The estimated common and private streams at IR-$k$ are $\hat{s}_{c,k}^{\mathrm{ID}}=g_{c,k}y_{k}^{\mathrm{ID}}$ and $\hat{s}_{k}^{\mathrm{ID}}=g_{c,k}(y_{k}^{\mathrm{ID}}-\mathbf{h}_k^H\mathbf{p}_c{s}_{c}^{\mathrm{ID}})$. The Mean Square Errors (MSEs) are respectively defined as
\vspace{-1mm}
\begin{equation}
\begin{aligned}
	&\varepsilon_{c,k}\triangleq\mathbb{E}\{|\hat{s}_{c,k}^{\mathrm{ID}}-s_c^{\mathrm{ID}}|^2\}=|g_{c,k}|^2T_{c,k}-2\Re\{g_{c,k}\mathbf{h}_{k}^H\mathbf{p}_{c}\}+1,\\
	&\varepsilon_{k}\triangleq\mathbb{E}\{|\hat{s}_{k}^{\mathrm{ID}}-s_{k}^{\mathrm{ID}}|^2\}=|g_{k}|^2T_{k}-2\Re\{g_{k}\mathbf{h}_{k}^H\mathbf{p}_{k}\}+1,
	\end{aligned}
	\vspace{-0.5mm}
\end{equation}
where $T_{c,k}\triangleq|\mathbf{h}_k^H\mathbf{p}_{c}|^2+\sum_{j\in\mathcal{K}}|\mathbf{h}_k^H\mathbf{p}_{j}|^2+1$ and $T_{k}\triangleq\sum_{j\in\mathcal{K}}|\mathbf{h}_k^H\mathbf{p}_{j}|^2+1$. By solving $\frac{\partial\varepsilon_{c,k}}{\partial g_{c,k}}=0$ and $\frac{\partial\varepsilon_{k}}{\partial g_{k}}=0$, we derive the optimal MMSE equalizers which are given by
\vspace{-0.5mm}
\begin{equation}
	g_{c,k}^{\textrm{MMSE}}=\mathbf{p}_{c}^H\mathbf{h}_k{T}_{c,k}^{-1},\,\,
	g_{k}^{\textrm{MMSE}}=\mathbf{p}_{k}^H\mathbf{h}_k{T}_{k}^{-1}.
	\vspace{-0.5mm}
\end{equation}
The Minimized MSEs (MMSEs) based on $g_{c,k}^{\mathrm{MMSE}}$ and $g_{k}^{\mathrm{MMSE}}$ are given by
\vspace{-1mm}
\begin{equation}
\begin{aligned}
&\varepsilon_{c,k}^{\textrm{MMSE}}\triangleq\min_{g_{c,k}} \varepsilon_{c,k} ={T}_{c,k}^{-1}(T_{c,k}-|\mathbf{h}_k^H\mathbf{p}_{c}|^2),\\
&\varepsilon_{k}^{\textrm{MMSE}}\triangleq\min_{g_{k}} \varepsilon_{k} ={T}_{k}^{-1}(T_{k}-|\mathbf{h}_k^H\mathbf{p}_{k}|^2).
\end{aligned}
\vspace{-1mm}
\end{equation}
The SINRs of decoding the intended streams are $\gamma_{c,k}={1}/{\varepsilon_{c,k}^{\textrm{MMSE}}}-1$ and $\gamma_{k}={1}/{\varepsilon_{k}^{\textrm{MMSE}}}-1$, respectively. The corresponding common rate and private rate of IR-$k$ are $R_{c,k}=\log_2(1+\gamma_{c,k})$ and $R_{k}=\log_2(1+\gamma_{k})$, respectively. The augmented WMSEs are
\vspace{-1mm}
\begin{equation}
\begin{aligned}
\xi_{c,k}=w_{c,k}\varepsilon_{c,k}-\log_2(w_{c,k}),\,\,
\xi_{k}=w_{k}\varepsilon_{k}-\log_2(w_{c}),
\end{aligned}
\vspace{-1mm}
\end{equation}
where $w_{c,k}$ and $w_{k}$ are the weights of the MSEs of IR-$k$. Note that the weights $w_{k},w_{c,k}$ of the MSEs are different from the weight $u_k$ allocated to the rate of each IR. By solving $\frac{\partial\xi_{c,k}}{\partial g_{c,k}}=0$ and $\frac{\partial\xi_{k}}{\partial g_{k}}=0$, the optimal equalizers are the same as MMSE equalizers. The corresponding optimal augmented WMSEs are 
\vspace{-1mm}
\begin{equation}
\label{eq: optimal WMSEs}
\begin{aligned}
&\xi_{c,k}(g_{c,k}^{\mathrm{MMSE}})=w_{c,k}\varepsilon_{c,k}^{\mathrm{MMSE}}-\log_2(w_{c,k}),\\
&\xi_{k}(g_{k}^{\mathrm{MMSE}})=w_{k}\varepsilon_{k}^{\mathrm{MMSE}}-\log_2(w_{c}).
\end{aligned}
\vspace{-1mm}
\end{equation}
By further solving $\frac{\partial\xi_{c,k}(g_{c,k}^{\mathrm{MMSE}})}{\partial w_{c,k}}=0$  and $\frac{\partial\xi_{k}(g_{k}^{\mathrm{MMSE}})}{\partial w_{k}}=0$, we obtain the optimal weights of the MMSEs as 
\vspace{-1mm}
\begin{equation}
\label{eq: optimal weights}
\begin{aligned} w_{c,k}^*=w_{c,k}^{\textrm{MMSE}}\triangleq(\varepsilon_{c,k}^{\textrm{MMSE}})^{-1},\,\, w_{k}^*=w_{k}^{\textrm{MMSE}}\triangleq(\varepsilon_{k}^{\textrm{MMSE}})^{-1}.
\end{aligned}
\vspace{-0.5mm}
\end{equation}
Substituting (\ref{eq: optimal weights}) into (\ref{eq: optimal WMSEs}), the Rate-WMMSE relationships are established as
\vspace{-1mm}
\begin{equation}
\label{eq: Rate wmmse relation}
\begin{aligned}
   \xi_{c,k}^{\textrm{MMSE}}(\mathbf{P})\triangleq 1-R_{c,k}(\mathbf{P}),\,\,
   \xi_{k}^{\textrm{MMSE}}(\mathbf{P})\triangleq 1-R_{k}(\mathbf{P}).
\end{aligned}
\vspace{-1mm}
\end{equation}

\par Based on the Rate-WMMSE relationship in (\ref{eq: Rate wmmse relation}), the original  problem (\ref{eq: RE prob}) is reformulated as
\vspace{-2mm}
 \begin{subequations}
	\label{eq: RE prob WMMSE}
	\begin{align}
	\min_{\mathbf{{P}},\mathbf{{F}},\mathbf{x},\mathbf{w},\mathbf{g}}& \sum_{k\in\mathcal{K}}u_k(X_k+\xi_{k}(\mathbf{P})) \\
	\mbox{s.t.}\quad
	& \sum_{k\in\mathcal{K}}X_{k}+1\geq\xi_{c,k}(\mathbf{P}), ,\forall k\in\mathcal{K} \label{const: common}\\
	&\sum_{j\in\mathcal{J}}Q_j\geq E^{th}\label{const: energy}\\
	&	\text{tr}(\mathbf{P}\mathbf{P}^{H})+\mathrm{tr}(\mathbf{F}\mathbf{F}^{H})\leq P_{t} \label{const: power}\\
	&\mathbf{x}\leq\mathbf{0} \label{const: WMSE}
		\vspace{-2mm}
	\end{align}	
\end{subequations} 
where $X_k$ is the transformed WMSE that corresponds to the common rate allocated to user-$k$. $\mathbf{x}=[X_1,\ldots,X_K]$ is the WMSE vector. $\mathbf{w}=[w_1,\ldots,w_K,w_{1,c},\ldots,w_{K,c}]$ is the vector of all the MSE weights. $\mathbf{g}=[g_1,\ldots,g_K,g_{1,c},\ldots,g_{K,c}]$ is the vector containing all the equalizers. 

Different from the WMMSE transformation introduced in \cite{RS2016hamdi} for the WSR maximization problem of RS in WIT, problem (\ref{eq: RE prob WMMSE}) is still non-convex with respect to $\{\mathbf{{P}},\mathbf{{F}},\mathbf{x}\}$ due to the conflicting  harvested energy constraint (\ref{const: energy}) and power consumption constraint (\ref{const: power}). 
To solve the problem, we  further carry out the first-order Taylor expansion to the harvested energy at each user. Based on the  first-order lower bound of $|\mathbf{g}_{j}^{H}\mathbf{p}_{k}|^{2}$ at a given point $\mathbf{p}_{k}^{[t]}$, which is given by
\vspace{-1mm}
\begin{equation}
\begin{aligned}
|\mathbf{g}_{j}^{H}\mathbf{p}_{k}|^{2}
&\geq 2\mathrm{Re}\left((\mathbf{p}_{k}^{[t]})^{H}\mathbf{g}_{j}\mathbf{g}_{j}^{H}\mathbf{p}_{k}\right)-|\mathbf{g}_{j}^{H}\mathbf{p}_{k}^{[t]}|^{2}\\
&\triangleq\Phi^{[t]}(\mathbf{p}_{k},\mathbf{g}_{j}),
\end{aligned}
\vspace{-1mm}
\end{equation}
constraint (\ref{const: energy}) becomes
\vspace{-1mm}
\begin{equation}
\label{const: energy app}
\Phi^{[t]}(\mathbf{p}_{c},\mathbf{g}_{j})+
\sum_{k\in\mathcal{K}}\Phi^{[t]}(\mathbf{p}_{k},\mathbf{g}_{j})+\sum_{j'\in\mathcal{J}}\Phi^{[t]}(\mathbf{f}_{j'},\mathbf{g}_{j})\geq E_j^{th}.
\vspace{-1mm}
\end{equation}
Hence, problem (\ref{eq: RE prob WMMSE}) is approximated by
\vspace{-1mm}
 \begin{subequations}
	\label{eq: RE prob WMMSE SCA}
	\begin{align}
	\min_{\mathbf{{P}},\mathbf{{F}},\mathbf{x},\mathbf{w},\mathbf{g}}& \sum_{k\in\mathcal{K}}u_k(X_k+\xi_{k}(\mathbf{P}))  \\
	\mbox{s.t.}\quad
	& \textrm{(\ref{const: common}), 
		 (\ref{const: energy}), (\ref{const: WMSE}), (\ref{const: energy app})}.
		\vspace{-2mm}
	\end{align}	
\end{subequations}
With fixed $\{\mathbf{w},\mathbf{g}\}$, problem (\ref{eq: RE prob WMMSE SCA}) is now a Quadratic Constrained Quadratic Programming (QCQP) at the point $\mathbf{P}^{[t]},\mathbf{F}^{[t]}$, which can be solved efficiently by standard convex optimization methods, e.g., interior point method. The convexity of problem (\ref{eq: RE prob WMMSE SCA}) in each iteration $[t]$ motivates us to use the SCA-based algorithm illustrated in Algorithm \ref{SCA algorithm} to update $\mathbf{P}^{[t]},\mathbf{F}^{[t]}$ iteratively. In each iteration, problem (\ref{eq: RE prob WMMSE SCA}) is solved and the variables  $\mathbf{P}^{[t]},\mathbf{F}^{[t]},\mathbf{x}$ are updated using the corresponding optimized solution. $\mathrm{WMMSE}^{[t]}=\sum_{k\in\mathcal{K}}u_k(X_k^{[t]}+\xi_{k}(\mathbf{P}^{[t]})) $ is the WMMSE calculated based on the updated $(\mathbf{P}^{[t]},\mathbf{x}^{[t]})$ at iteration $[t]$. $\epsilon$ is the tolerance of the algorithm.
\begin{algorithm}[t!]
	\textbf{Input}: $t\leftarrow0$, $\mathbf{P}^{[t]}$, $\mathrm{F}^{[t]}$, $\mathbf{w}, \mathbf{g}$\;
	\Repeat{$|\mathrm{WMMSE}^{[t]}-\mathrm{WMMSE}^{[t-1]}|\leq \epsilon$}{
		$t\leftarrow t+1$\;
		$\mathbf{P}^{[t-1]}\leftarrow \mathbf{P}^{[t]}$\;
		$\mathbf{F}^{[t-1]}\leftarrow \mathbf{F}^{[t]}$\;

		update $(\mathbf{P}^{[t]},\mathbf{F}^{[t]},\mathbf{x}^{[t]})$ by solving problem (\ref{eq: RE prob WMMSE SCA}) using the fixed $\mathbf{w}, \mathbf{g}$ and updated $\mathbf{P}^{[t-1]}, \mathbf{F}^{[t-1]}$\;		
		update 	$\mathrm{WMMSE}^{[t]}$ using	$(\mathbf{P}^{[t]},\mathbf{x}^{[t]})$
	}	
	\caption{SCA-based algorithm}
	\label{SCA algorithm}					
\end{algorithm}
	
Algorithm \ref{SCA algorithm} derives the near optimal solution of problem  (\ref{eq: RE prob WMMSE SCA}) when $\{\mathbf{w},\mathbf{g}\}$  are fixed. According to the KKT condition of problem (\ref{eq: RE prob WMMSE SCA}), it is easy to verify that  the MMSE equalizer $\mathbf{g}^{\textrm{MMSE}}=\{g_{k}^{\textrm{MMSE}},g_{c,k}^{\textrm{MMSE}}|\forall k\in\mathcal{K}\}$ is the optimal solution of $\mathbf{g}$ with fixed $\{\mathbf{{P}},\mathbf{{F}},\mathbf{x},\mathbf{w}\}$.
The MMSE weight $\mathbf{w}^{\textrm{MMSE}}=\{w_{k}^{\textrm{MMSE}},w_{c,k}^{\textrm{MMSE}}|\forall k\in\mathcal{K}\}$ is the optimal solution of $\mathbf{w}$ with fixed $\{\mathbf{{P}},\mathbf{{F}},\mathbf{x},\mathbf{g}\}$. The partial convexity of problem (\ref{eq: RE prob WMMSE SCA}) motivates us to use the Alternating Optimization (AO) algorithm to solve the problem. The details of the algorithm is as shown in Algorithm \ref{WMMSE algorithm}. 
 $\mathrm{WSR}^{[n]}=\sum_{k\in\mathcal{K}}u_k\left(C_k^{[n]}+R_{k}(\mathbf{P}^{[n]})\right) $ is the WSR calculated based on the updated $(\mathbf{P}^{[n]},\mathbf{c}^{[n]})$. 
\begin{algorithm}[t!]
	\textbf{Initialize}: $n\leftarrow0$, $\mathbf{P}^{[n]}$, $\mathbf{F}^{[n]}$, $\mathrm{WSR}^{[n]}$\;
	\Repeat{$|\mathrm{WSR}^{[n]}-\mathrm{WSR}^{[n-1]}|\leq \epsilon$}{
		$n\leftarrow n+1$\;
		$\mathbf{P}^{[n-1]}\leftarrow \mathbf{P}^{[n]}$\; $\mathbf{F}^{[n-1]}\leftarrow \mathbf{F}^{[n]}$\;
		$\mathrm{WSR}^{[n-1]}\leftarrow \mathrm{WSR}^{[n]}$\;
		$\mathbf{w}^*\leftarrow\mathbf{w}^{\mathrm{MMSE}}(\mathbf{P}^{[n-1]})$\; $\mathbf{g}^*\leftarrow\mathbf{g}^{\mathrm{MMSE}}(\mathbf{P}^{[n-1]})$\;
	
		update $(\mathbf{P}^{[n]},\mathbf{F}^{[n]},\mathbf{x}^{[n]})$ by  using \textbf{Algorithm \ref{SCA algorithm}} with  the updated $\mathbf{w}^*, \mathbf{g}^*,\mathbf{P}^{[n-1]},\mathbf{F}^{[n-1]}$\;
		$\mathbf{c}^{[n]}\leftarrow -\mathbf{x}^{[n]}$\;
	   update 	$\mathrm{WSR}^{[n]}$ using	$(\mathbf{P}^{[n]},\mathbf{c}^{[n]})$
	}	
	\caption{WMMSE-based AO algorithm}
	\label{WMMSE algorithm}		
\end{algorithm}

\par \textit{Convergence Analysis}:
For a given $\{\mathbf{w},\mathbf{g}\}$, the convergence of the proposed SCA-based algorithm is guaranteed. The solution $\{\mathbf{P}^{[t]},\mathbf{F}^{[t]}\}$ at iteration $[t]$ is also a feasible solution at iteration  $[t+1]$. Hence, $\mathrm{WMMSE}$ is monotonically decreasing  and it is bounded below by the sum power constraint. 
The proposed WMMSE-based AO algorithm is guaranteed to converge as well since the  solution $\{\mathbf{w}^*, \mathbf{g}^*,\mathbf{P}^{[n]},\mathbf{F}^{[n]}\}$ at iteration $[n]$ is also a feasible solution at iteration  $[n+1]$. Therefore, the objective $\mathrm{WSR}$ increases iteratively and it is bounded above due to the sum power constraint of BS. 


%
%
%
%

%
%
%
%
%
%
%
\vspace{-1mm}
\section{Numerical Results}
\vspace{-1mm}
\label{sec: simulation}
\par In this section, we evaluate the performance of the proposed RS-assisted strategy by comparing with MU--LP where each IR directly decodes the intended  stream and SC--SIC where the IR with stronger channel strength is required to decode the stream of the IR with weaker channel strength.  Of particular interest, we ask ourselves how MU--LP, SC--SIC and RS strategies compare as the physical locations of IRs and ERs change.  
To answer this question, we consider the situation when there are two IRs and one ER in the system. The channel angle between the ER and each IR is varied while different  channel angle and channel strength disparity between the IRs are investigated. 

\par To consider the large-scale path loss, the channels of IRs and ERs are constructed as $\mathbf{h}_k=d_h^{-\frac{3}{2}}\widetilde{\mathbf{h}}_k, \forall k\in\mathcal{K}$ and $\mathbf{g}_j=d_g^{-\frac{3}{2}}\widetilde{\mathbf{g}}_j, \forall j\in\mathcal{J}$ as in \cite{MMSESWIPT2016Song}.  $d_h$ and $d_g$ are set to 10 meters. Following the two-user deployments in \cite{mao2017rate}, we consider specific channel realizations for the IRs given by  
$
\widetilde{\mathbf{h}}_1=\left[1, 1, 1, 1\right]^H$,
$\widetilde{\mathbf{h}}_2=\gamma\times[
1, e^{j\theta},e^{j2\theta}, e^{j3\theta}]^H$.
$\gamma$ and  $\theta$ control the channel strength disparity and channel angle between IRs, respectively. 
The channel realization of ER is $\widetilde{\mathbf{g}}_1=[
1, e^{j\beta},e^{j2\beta}, e^{j3\beta}]^H$.  When $\beta=0$, the location of ER coincides with that of IR-1. When $\beta=\theta$, ER is aligned with IR-2. The transmit power is fixed to $P_t=10$ dBm. The noise power is assumed to be equal to $-30$ dBm at each IR.

 \begin{figure}[t!]
\vspace{-5mm}
	\small
	\centering
	\includegraphics[width=3.0in]{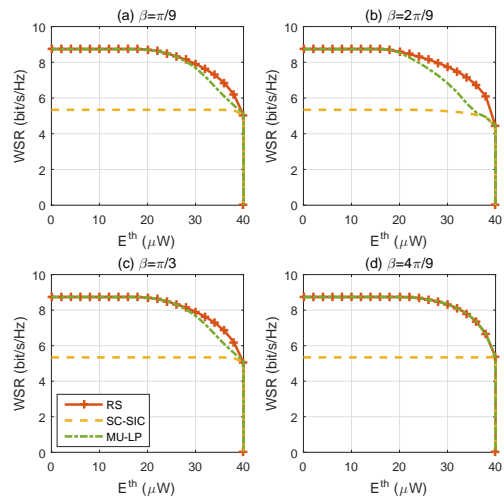}%
\vspace{-2mm}
	\caption{Rate-energy region comparison  of different strategies with $\gamma=1$, $\theta=4\pi/9$.}
	\label{fig: rateEnergy bias1theta80}
\vspace{-5mm}
\end{figure}

 \par Fig. \ref{fig: rateEnergy bias1theta80} shows the rate-energy tradeoff when $\gamma=1$, $\theta=4\pi/9$. The weight pair given to IRs is  $u_1=u_2=1$. As $E^{th}$ approaches the largest achievable value (30 $\mu$W--40 $\mu$W), the performance benefit of RS over MU--LP and SC--SIC becomes obvious. 
To  understand the rationale behind the WSR gap  of RS over MU--LP and SC--SIC, we specify the power allocated to each precoder  when $\theta=2\pi/9$, $E^{th}=35$ $\mu$W (Fig. \ref{fig: rateEnergy bias1theta80}(b)) in Table \ref{tab: power allocation}. $P_c=\mathrm{tr}(\mathbf{p}_c\mathbf{p}_c^{H})$, $P_1=\mathrm{tr}(\mathbf{p}_1\mathbf{p}_1^{H})$, $P_2=\mathrm{tr}(\mathbf{p}_2\mathbf{p}_2^{H})$   and
$P_{\textrm{ER}}=\mathrm{tr}(\mathbf{f}_1\mathbf{f}_1^{H})$. 
 We observe that MU--LP  allocates dedicated power for the precoder of ER $\mathbf{f}_1$ at the transmitter  in order to meet the ER energy constraint. However, as $\mathbf{f}_1$ is used for ER only, less power is allocated to transmit the data streams of IRs. The WSR of IRs deteriorates when more power is allocated to $\mathbf{f}_1$.  In comparison, as the common stream is required to be decoded by both users, RS is able to use the precoder of the common stream $\mathbf{p}_c$ to transmit both energy to the ER and information to the IRs. The energy harvested at the ER can be guaranteed while the WSR of IRs is not deteriorated. Same observations are obtained when we investigate other scenarios where the ER does not coincide with both IRs. 
\begin{table}[h!]
	\caption{Power allocation in  Fig. \ref{fig: rateEnergy bias1theta80}(b), $E^{th}=35$ $\mu$W}
	\label{tab: power allocation}
	\centering
	\begin{tabular}{|l|l|l|l|l|l|l|}
		\hline
		& WSR  (bit/s/Hz) & $P_c$ (W)     & $P_1$  (W)   & $P_2$  (W)   & $P_{\textrm{ER}}$ (W)\\ \hline
		\textbf{RS}      &  6.9598         &  0.0074 &  0.0013 & 0.0013  &   0    \\ \hline
		\textbf{MU--LP}  &  5.3265        &    -      &    0.0017   &    0.0017       &  0.0066    \\ \hline
		\textbf{SC--SIC} &  5.1086         &    -       &    0.0015      &     0.0085      &   0    \\ \hline
	\end{tabular}
\vspace{-3mm}
\end{table}

\begin{figure}[t!]
	\vspace{-4mm}
	\small
	\centering
	\includegraphics[width=3.0in]{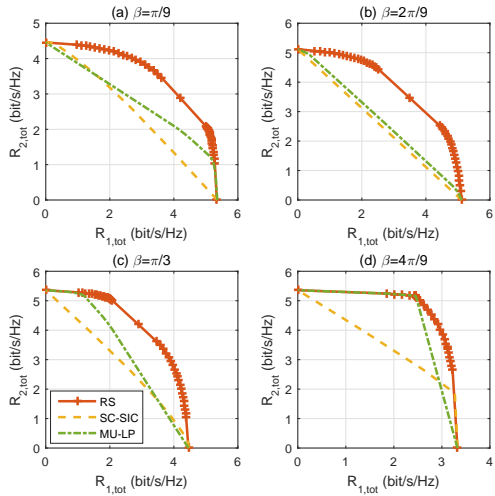}%
	\vspace{-2mm}
	\caption{IR rate region comparison of different strategies with $\gamma=1$, $\theta=4\pi/9$, $E^{th}=35$ $\mu W$.}
	\label{fig: rateRegion Eth35bias1theta80}
	\vspace{-4mm}
\end{figure}

\begin{figure}[t!]
	\vspace{-4mm}
	\small
	\centering
	\includegraphics[width=3in]{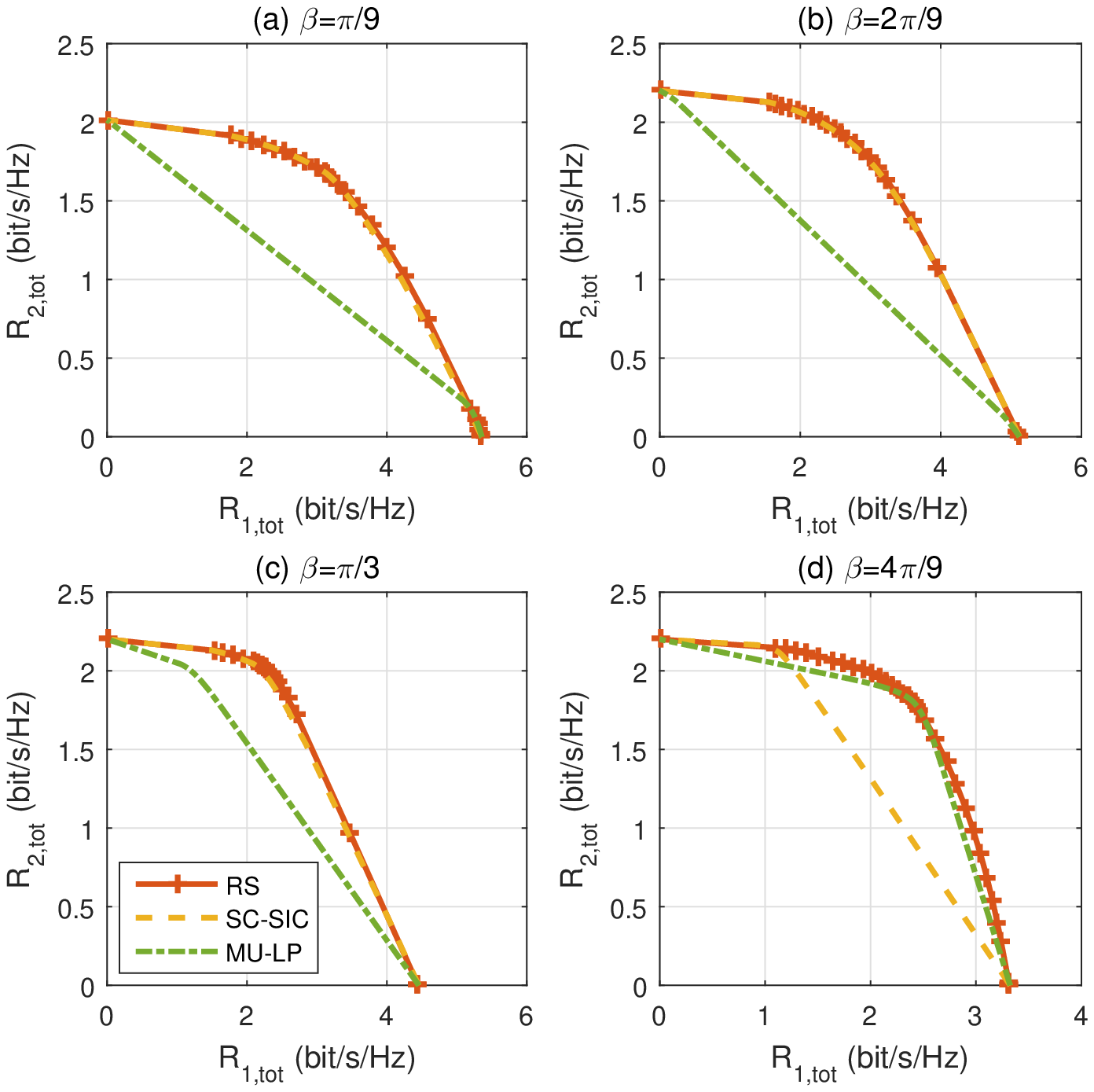}%
	\vspace{-2mm}
	\caption{IR rate region comparison of different strategies with $\gamma=0.3$, $\theta=\pi/3$, $E^{th}=35$ $\mu W$.}
	\label{fig: rateRegion Eth35bias03theta60}
	\vspace{-5mm}
\end{figure}

The two-IR rate regions when $\gamma=1, \theta=4\pi/9$ and $\gamma=0.3, \theta=\pi/3$ are illustrated in Fig. \ref{fig: rateRegion Eth35bias1theta80} and Fig. \ref{fig: rateRegion Eth35bias03theta60}, respectively. The boundary of the rate region is obtained by solving  Problem (\ref{eq: RE prob}) using various weight pairs for the two IRs. Following \cite{RS2016hamdi, mao2017rate}, the weight of IR-1 is fixed to 1 ($u_1=1$) while the weight of IR-2 is varied as $u_2=10^{[-3, -1,-0.95,\ldots,0.95,1,3]}$. 
There is an explicit IR rate region improvement of RS over MU--LP and SC--SIC in Fig. \ref{fig: rateRegion Eth35bias1theta80}. In the information-only transmission, it has been illustrated in \cite{mao2017rate} that the rate region of RS reduces to MU--LP when user channels are orthogonal. In comparison, RS reaps the benefit of the precoder for the introduced common stream in SWIPT to simultaneously transfer information and power to the IRs and ER. It is able to achieve a better rate region than MU--LP even when user channels are orthogonal. Comparing Fig. \ref{fig: rateRegion Eth35bias1theta80} and Fig. \ref{fig: rateRegion Eth35bias03theta60}, we observe that SC--SIC is more suited to the cases when the channel strength disparity of the IRs is large and the channel angle between the IRs is small. The proposed RS-assisted strategy bridges SC--SIC and MU--LP and is able to achieve a better IR rate region for a given harvested energy constraint of the ER.

\section{Conclusions}
\label{sec: conclusion}
To conclude, we propose an RS-assisted strategy in SWIPT with separated IRs and ERs. We investigate the WSR maximization of IRs under the harvested energy constraint of ERs and total transmit power constraint. A WMMSE and SCA-based algorithm is proposed to solve the problem efficiently. Numerical results show that the proposed RS-assisted strategy achieves a better rate-energy tradeoff for SWIPT compared with conventional MU--LP and SC--SIC strategies. This is contributed by the precoder of the common stream since it has the dual purpose of transmitting information to IRs and carrying power to the ERs  in SWIPT.
Therefore, we conclude that RS is a more 
powerful transmission scheme for downlink SWIPT.

\bibliographystyle{IEEEtran}
\vspace{-2mm}
\bibliography{reference}	
\vspace{-6mm}

\end{document}